# Onset of coherent attitude layers in a population of sports fans

Robert Hristovski[1], Natàlia Balagué[2], Josep Juaneda[2] and Pablo Vázquez[2]

[1]Faculty of Physical Education, Ss. Cyril and Methodius University – Skopje, Republic of Macedonia
[2]Institut Nacional dEducació Física (INEFC), Universitat de Barcelona, Spain

**Abstract**

The aim of this paper was to empirically investigate the behavior of fans, globally coupled to a common environmental source of information. The environmental stimuli were given in a form of referee's decisions list. The sample of fans had to respond on each stimulus by associating points signifying his/her own opinion, emotion and action that referee's decisions provoke. Data were fitted by the Brillouin function which was a solution of an adapted model of quantum statistical physics to social phenomena. Correlation and a principal component analysis were performed in order to detect any collective behavior of the social ensemble of fans. Results showed that fans behaved as a system subject to a phase transition where the neutral state in the opinion, emotional and action space has been destabilized and a new stable state of coherent attitudes was formed. The enhancement of fluctuations and the increase of social susceptibility (responsiveness) to referee's decisions were connected to the first few decisions. The subsequent reduction of values in these parameters signified the onset of coherent layering within the attitude space of the social ensemble of fans. In the space of opinions fan coherence was maximal as only one layer of coherence emerged. In the emotional and action spaces the number of coherent levels was 2 and 4 respectively. The principal component analysis revealed a strong collective behavior and a high degree of integration within and between the opinion, emotional and action spaces of the sample of fans. These results point to one possible way of how different proto-groups, violent and moderate, may be formed as a consequence of global coupling to a common source of information.

## Introduction

Coherent states appear very often in social systems of different kinds and cardinality. Due to its significance, with respect to the survival of the community, this phenomena has been present throughout the whole evolutionary scale, both in cell and animal communities, for example, in the form of orientation consolidation when escaping a predator, in seasonal migrations, as well as in human social groups (for instance the formation of pockets of coherent opinion holders). A striking property of the coherence is that it represents a qualitatively new macroscopically ordered state that emerges from a previous stochastic state of the system, under the influence of a certain non-specific parameter change.

One of the earliest attempts to predict qualitative changes in social systems was the cellular automata model, proposed by Shelling, (1969, 1971) for studying the ethnic clustering (segregation) phenomenon. Several mathematical models that explained qualitative changes in social systems, as a consequence of a nonspecific parameter change were proposed by several authors (e.g. Zeeman et al., 1976; Isnard and Zeeman, 1976; Weidlich. 1977; Renfrew and Poston, 1978). These models have found their plausible incarnation in the basic neural network and cellular automata concepts. This approach showed an exquisite explanatory power (by means of computer simulations) of macroscopic effects of the inter-individual social interactions (Nowak et al., 1990; Nowak and Hegselmann, 1993; Nowak and Latane, 1994; Latane, 1996; Nowak and Vallacher, 1998; Nowak et al. 1998; Vallacher et al., 2010; Vallacher and Jackson, 2009) such as emergence of clustering, attitude polarization, ideologies, beliefs etc. Some interesting results have been obtained within the two - parameter class of probabilistic cellular automata simulations known as the nonlinear voter model (Gilpin, 1975; May and Leonard, 1975; Cox and Durrett, 1991; Silvertown et al. 1992; Liggett, 1994; Tainaka, 1995) according to which interspecies competition leads to the onset of different macroscopic social effects such as: coexistence, ergodicity, and clustering or phase separation.

All these models share the common characteristic of being local models. In other words, the coherence within these models emerges as a result of local interactions between social elements. However, coherent states may also emerge when social elements possess a common coupling with environmental events but are not mutually coupled. Complex dynamical systems, particularly fan groups, start forming from some initial state, evolve and stabilize through mutual interactions and eventually they may dissolve. This initial stadium, in which members of some social ensemble are still not interacting but form coherent layers, we define as a state of a proto-group, i.e. a group to-be-formed. This subtle state, between the phases of a networked, full blown group and absolute attitude de-coherence can be considered as a critical nucleus, a precursor, which, depending on constraints, may evolve into a group or disintegrate back to a population of independent individuals. The aim of this study was to explain one possible way in which such initial stadiums of complex sociological networks may be formed and that is the non-local emergence of coherent opinions, emotions and action preferences as a consequence of a global coupling of a social ensemble with an external source of information.

A model, developed some time ago (Hristovski, 1997), predicts onset of non - local coherence in social ensembles that consist of locally non - interacting individuals, coupled globally and unidirectionally only to the events in their environment. The network topology of such agent-environment systems may be defined as a star topology (see Fig.1) where the central node activity represents an environmental event (e.g. a referee's decision – $B_0$) impinging on peripheral nodes, e.g. team supporters.

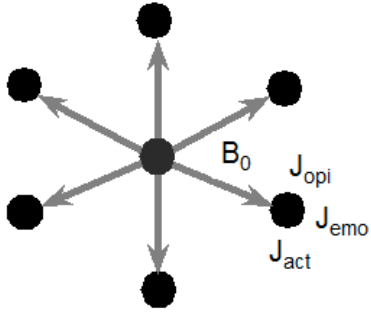

**Figure 1.** A star network of a referee-fans system. Team supporters are given as black circles designating their common sign of coupling, e.g.+1 or -1, with the referee's decision. Each fan may have different personal constraints J in the opinion, emotional and action space.

This is a rather different class of social network interaction compared to the above mentioned models. Due to the non - interacting nature of the peripheral nodes (e.g. fans) the model has exact solutions that can be easily experimentally evaluated. The model predicts the final states of some social phenomena such as the formation of proto - groups, which have much in common with the local models mentioned above. Note that although there is no immediate and explicit interaction (information exchange) between the members of the social ensemble of fans, they are nonetheless bound together through their common characteristic. They are supporters of the same team and have identical or at least strongly correlated values, i.e. identification with the team. In this aspect we can speak of a non-local or an implicit coupling, within the social ensemble of explicitly non-interacting fans.

Obviously, due to the interaction of the environmental constraint, i.e. a referee's decision, and their common personal constraints, i.e. identification with their favorite team, a new set of common properties may be induced in the group in a form of coherent opinions, emotions and preferable actions. These properties may form an additional set of constraints, juxtaposed to and interacting with the previous ones, which together could form another, more complex initial stadium that may initiate the evolution of such proto-group.

The exact solution of the model is a Brillouin function that describes the behavior of a social ensemble under environmental influences B:

$$\langle \sigma_j \rangle = \frac{2J+1}{2J} cth \frac{2J+1}{2J}(\beta gJmB) - \frac{1}{2J} cth \frac{1}{2J}(\beta gJmB) \quad (1)$$

Here J has the role of a personal constraint in the attitudinal space (see Fig.1). More precisely, it represents the extremity that could be attained by the fans in some of the attitudinal components such as opinion, emotional and action responses. The number of levels that the social system may attain is $2J+1$. g is a splitting Lande factor. Due to the bipolarity, there should be the same number of states on the right and left side of the zero. β and m are free parameters of the model.

$$\langle U_j \rangle = gJm \langle \sigma_j \rangle \quad (2)$$

is the expected attitudinal charge per subject where gJm gives the asymptotic value of coherent layers which have been formed. The social susceptibility, i.e. responsiveness, to the external stimuli per subject is defined as:

$$\chi = \frac{\langle U_J \rangle}{B} \quad (3)$$

The following explicit predictions arise within the framework of this model:

1. If the social ensemble is homogeneous in the coupling with the environment, then, an attitudinal coherence will arise for some value of the external stimulation $B_0$;
2. If the social ensemble is not homogeneous (simultaneously contains different coupling signs with the environmental events) then, an attitudinal polarization will arise for some value of the external stimulation $B_0$;
3. If the social ensemble is homogeneous with respect to the internal personal constraint J, then, a single layered attitudinal coherence will arise for some value of the external stimulation $B_0$;
4. If the social ensemble is not homogeneous with respect to the personal constraint J, then, a multilayered attitudinal coherence, proportional to the number of different values of J will arise for some value of the external stimulation $B_0$;
5. If the social ensemble is not homogeneous with respect to both - the common coupling with the environmental events and personal constraint (J), then, for some value of the external stimulation $B_0$, a polarized multilayered coherence will arise with a maximum of $2J+1$ layers.

In order to test some of these predictions, we have performed the following evaluation procedure. We have treated a sample of participants homogenously coupled with the environmental events. We sought to determine a number of coherent layers, i.e. a number of different J constraints, present in the sample of fans.

## Method

### Participants

The sample of 97 participants ($N_{male} = 73$; $N_{female} = 24$), aged 20 - 24, was randomly extracted from the population of students at the Faculty for Physical Culture in Skopje. They were not aware of the purpose of the experiment.

**List of Environmental Stimuli and Questionnaires**

One list of environmental stimuli (referee's decisions) and 3 questionnaires were designed to serve the purpose of the experiment.

1. The list of environmental stimuli ($B_0$) contained 12 stimuli of incorrect decisions concerning their favorite team. The list is with the first author. Because of spatial constraints we were not able to present it in detail. The list started as follows:
"*Imagine that you are watching on TV a football match in which your favorite team or the national team of your country is participating. The referee behaves in the following way:*"
Then the $B_0$ list of 12 referee decisions was presented to the test subjects. The 12 decisions covered the 2 halves of the match and they were defined and so explained to the participants as being of potential moderate influence on the match, for example decisions on fouls in the middle field.
The questionnaires contained $(2J+1)/2$ nonzero levels due to the fact that the sample belonged to a population of fans of a common team. Hence, the responses of the fans were expected to be aggregated on one side of the scale, i.e. to have a common orientation.

2. The questionnaire which assessed individual opinion change was based on scaling the subjects' judgments of the referee's decisions given in the foregoing list. The offered judgments were formed as probabilistic statements of the participants in accordance with the motives of the referee's decisions.
The questionnaire stated as follows:
"*Assess the behavior of the referee given on the list $B_0$ taking into account his previous decisions on the match. You have to associate a certain number of points on each of his decisions on the $B_0$ list.*" The list of responses containing the opinion responses of the participants stated as follows:
*0 - I don't have any opinion about his behavior, since he has not made any judgment yet;*
*1 - Something between 0 and 2*
*2 - His incorrect judgment is probably accidental;*
*3 - Something between 2 and 4*
*4 - It is rather likely that his incorrect judgments are not accidental;*
*5 - Something between 4 and 6*
*6 - It is almost sure that his incorrect judgments are not accidental and that he has something personally against my favorite team;*
*7 - Something between 6 and 8*
*8 - There is no doubt that his incorrect judgments are not accidental and that he has something personally against my favorite team;*

3. The questionnaire, assessing the individual emotional responses, was based on the scaling of the emotional states of the participants with respect to the environmental stimuli, given in list $B_0$. The offered emotional responses were formed as emotionally charged statements of the subjects with respect to the referee's decisions.

The questionnaire stated as follows:
"*Choose the emotion you have about each decision of the referee, taking in account his previous decisions during the match, by associating a certain number of points on the $B_0$ list*". The list of emotional responses of participants stated as follows:
*0 - I don't feel any emotion for the referee (neither positive nor negative);*
*1 - Something between 0 and 2;*
*2 - I don't like his behavior. He provokes a tiny anger in me;*
*3 - Something between 2 and 4;*
*4 - I don't like his behavior at all. He provokes a strong anger in me;*
*5 - Something between 4 and 6;*
*6- I can't withstand watching his behavior. He makes me very furious;*
*7- Something between 6 and 8;*
*8 - His behavior is disgusting. He provokes an unbearable fury in me.*

4. The questionnaire, assessing the individual preferred action responses, was based on scaling the action states that participants could attain with respect to the environmental stimuli (referee's decisions) given in list $B_0$. The offered action responses were formed as action charged statements of the subjects in accordance with the referee's decisions.
The questionnaire stated as follows: "*Write down what type of action provokes the referee's behavior given on the list $B_0$, taking into account his previous decisions on the match, so that you have to associate a certain number of points to each of his decisions on the $B_0$ list*". The list of action responses of participants stated as follows:
*0 - I would do nothing, since I don't know the referee;*
*1 - Something between 0 and 2;*
*2- His decisions have to be checked by the referee association;*
*3 - Something between 2 and 4;*
*4 - He has to be excommunicated from the duty;*
*5 - Something between 4 and 6;*
*6 - In addition to the excommunication I would like to have him in my hands for a while;*
*7 - Something between 6 and 8;*
*8 - In addition to the excommunication I would beat him to exhaustion.*

**Procedure**

The total sample was divided in 4 groups of approximately 24 participants. Participants were seated at distances that prevented cribbing (interactions among the subjects). The introduction text from every of the foregoing stimuli lists and scales was verbally explained and if needed some examples were given, in order to avoid misunderstandings about the demands of the experiment. Before each distribution of the opinion, emotion and action lists of responses, respectively, the lists of referee's decisions $B_0$ were distributed and the demands were explained. The experiment was conducted in the following order:

1. Distribution of the scales containing opinion responses ➞ explaining the experimental demands by the experimenter ➞ participants' responses to the stimuli ➞ collection of the responses done on the $B_0$ list;

2. Distribution of the scales containing emotional responses → explaining the experimental demands by the experimenter → participants' responses to the stimuli → collection of the responses done on the $B_0$ list;
3. Distribution of the scales containing action responses → explaining the experimental demands by the experimenter → participants' responses to the stimuli → collection of the responses done on the $B_0$ list;

### Data Analysis

Because of the analytical results and predictions of the model, for its evaluation we conducted standard statistical procedures. We estimated the free parameters m and β by applying a non-linear regression, by taking $g = 2$ and $J_{max} = 15/2$. The Quasi - Newton method was used for the determination of the parameter estimation convergence. The goodness of fit was tested by the amount of explained variances. The fit among the attitudinal evolutions in the opinion, emotional and the action space for different values of J were obtained by a linear regression procedure. A principal component analysis using the Guttman-Kaiser criterion for the determination of the number of significant components was applied to the twelve B dependent statistical averages of opinion, emotional and action attitudes (in total 7).

## Results

### Attitudinal Evolution in the Opinion, Emotional and Action Space

The non-linear regression (Equations 1, 2) showed highly significant relationships among the model predictions of the attitudinal evolutions and experimental data. The discreteness of the scale of attitudes does not enable a clear insight into the regions of maximal agglomeration of data. Look at Figure 5 and the text therein for help. In the opinion space (see Figure 2), the regression analysis revealed one layer of coherence (J = 7.5, N=97) and a high goodness of fit ($R^2$=.91, p<.001).

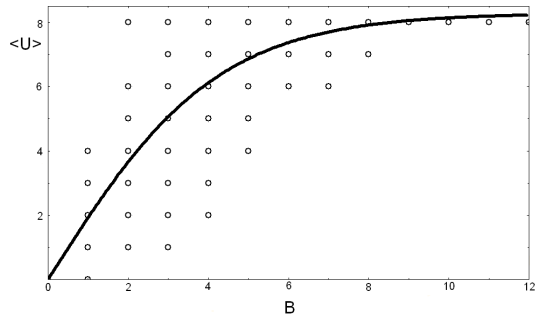

**Figure 2.** - Regression curve for the expected value of $<U>_{opinion}$. $J = 7.5$; $g = 2$; $m = 0.552$; $β = 0.1$. $R^2 = .905$; $p < .001$. One layer of coherence is formed for $B > 6$.

In the emotional space (Figure 3) two layers of coherence ($J_1 = 7.5$, $N = 71$ and $J_2 = 6.5$, $N = 26$) and highly significant relationships among the theoretical curves and data were found ($R^2_1 = .90$, $p < .001$ ; $R^2_2 = .88$, $p < .001$).

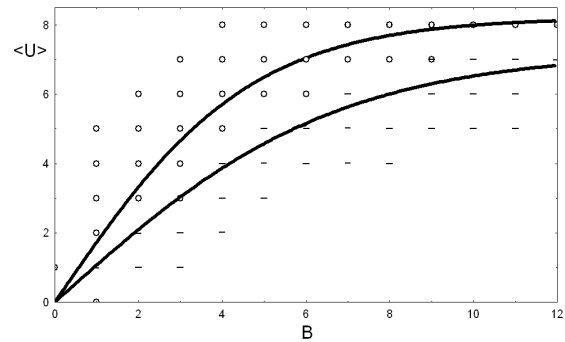

**Figure 3.** - Regression curves for $<U>_{emotional}$. $J_1 = 7.5$: $g = 2$; $m_1 = .547$; $β_1 = .094$. $R^2_1 = .897$; $p_1 < .001$. and $J_2 = 6.5$; $g = 2$; $m_2 = .515$; $β_2 = .073$. $R^2_2 = .884$; $p_2 < .001$. Two layers of coherence are formed: for ($J_1 = 7.5$) coherence emerged at $B > 6$ and for ($J_2 = 6.5$) at $B > 9$.

High goodness of fit was obtained in the action space (Figure 4) as well, where 4 coherent layers emerged ($J_1 = 7.5$, $N = 24$; $J_2 = 5.5$, $N = 18$; $J_3 = 4.5$, $N = 12$; and $J_4 = 3.5$, $N = 43$). The proportions of explained variance were respectively ($R^2_1 = .91$, p< .001; $R^2_2 = .93$, p< .001; $R^2_3 = .92$, p< .001 ; $R^2_4 = .80$, p < .001.

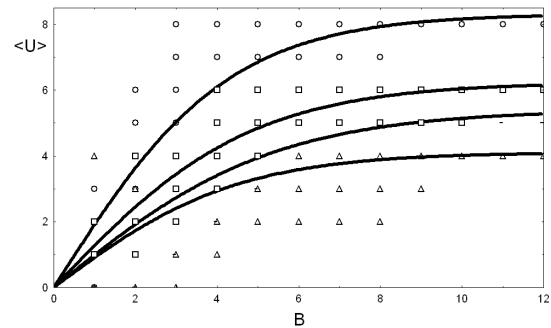

**Figure 4** - Regression curves for the expected value of $<U>_{action}$. $J_1 = 7.5$; $g = 2$; $m_1 = .554$; $β_1 = .099$; $R^2_1 = .908$; $p_1 < .001$; $J_2 = 5.5$; $g = 2$; $m_2 = .567$; $β_2 = .088$; $R^2_2 = .929$; $p_2 < .001$. $J_3 = 4.5$; $g = 2$; $m_3 = .531$; $β_3 = .139$. $R^2_3 = .918$; $p_3 < .001$. $J_4 = 3.5$; $g = 2$; $m_4 = .526$; $β_4 = .19$. $R^2_4 = .798$; $p_4 < .001$. Four layers of coherence are formed: For $J_1$, $J_2$ and $J_3$ the coherence arises at $B > 6$ and for $J_4$ at $B > 5$.

The residuals were symmetrically distributed around the expected values, which is another corroboration of the high goodness of fit among the theoretical predictions and the experimental data. From Figures 2 - 4 it can be seen that in most cases (using the criterion $J ≤ <U>$), the coherence emerged after the 5th and 6th stimulus ($B > 5$ and $B > 6$), except in the case $J = 7.5$ of the emotional space where the coherence emerged for $B > 9$. To get insight in the semantic charge (content) of the expected charge values <U>, see the see the Method part.

**Attitudinal Susceptibilities and Fluctuations in the Opinion, Emotional and Action Space**

The responsiveness and fluctuations did not show normal distribution densities. The sample responsiveness (susceptibility) attained its maximal values at median values of B=2 and fluctuations at median values of B = 3. The histograms in Figures 5 a and b, show the distributions of data from the opinion space for B = 3 and B = 8 respectively. For B=3 the opinion variability was maximal, however it converged to extremely low variability e.g. B=8. For higher values the variability vanished and coherence was formed.

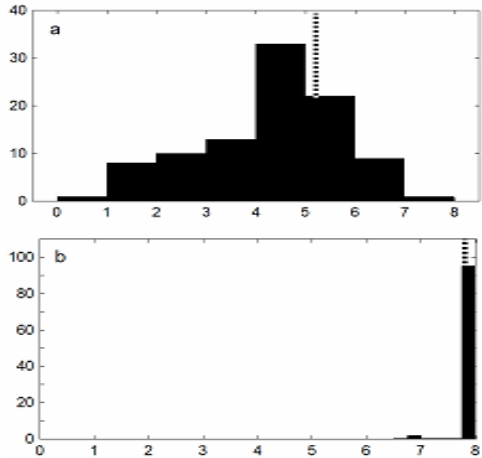

Figure 5. Histograms of data distribution in the opinion space: a. B=3 b. B=8. Dotted vertical lines signify the position of the fitted curve (compare with Fig.2). Vertical axis = number of observations; horizontal axis = upper boundaries (x ≤ boundary).

Figure 6 shows that the interval 0 < B < 4 is characterized by enhanced variability, i.e. fluctuations, and maximal susceptibility. After this interval, these sample characteristics started to slowly decrease which signals the formation of a new stable state and a condensation of attitudes toward their saturation values. Note that in contrast to fluctuations, the susceptibility does not converge to the zero value even for large B values. This is a consequence of the susceptibility definition (see Equation 3). If one considers the differential susceptibility that takes into account the change of <U> versus changes in B, than the point to point estimate of susceptibilities would show equal profile with fluctuations for large B values.

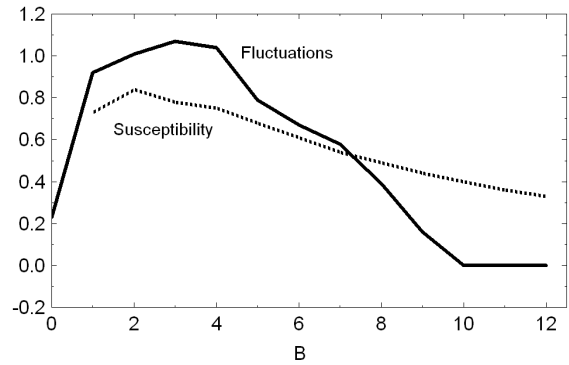

**Figure 6.** A typical profile of point to point estimate of susceptibilities from the data (dotted line -------- ) and fluctuation estimates from the data ( bold line).

## Attitudinal Patterns

We use the term 'attitudinal patterns' for the coordination (measured as a correlation) among the expected charge values (Equations 1 and 2) of the opinion, emotional and action components during the attitudinal evolution, arising under the influence of the environmental stimuli B.

Table 1 shows the high significance of the correlation among the attitudinal components.

| Correlation Pairs | $J_E$ | $R^2$ | p |
|---|---|---|---|
| $<U_O>$ vs. $<U_E>$ $J_O = 7.5$ | 7.5 | .996 | <.001 |
| | 6.5 | .941 | <.001 |
| | $J_A$ | | |
| $<U_O>$ vs. $<U_A>$ $J_O = 7.5$ | 7.5 | .999 | <.001 |
| | 5.5 | .993 | <.001 |
| | 4.5 | .981 | <.001 |
| | 3.5 | .996 | <.001 |
| | $J_A$ | | |
| $<U_E>$ vs. $<U_A>$ $J_E = 6.5$ | 7.5 | .939 | <.001 |
| | 5.5 | .962 | <.001 |
| | 4.5 | .953 | <.001 |
| | 3.5 | .960 | <.001 |
| | $J_A$ | | |
| $<U_E>$ vs. $<U_A>$ $J_E = 7.5$ | 7.5 | .994 | <.001 |
| | 5.5 | .998 | <.001 |
| | 4.5 | .988 | <.001 |
| | 3.5 | .995 | <.001 |

**Table 1.** Correlations among expected values in the opinion ($<U_C>$), emotional ($<U_E>$) and action ($<U_A>$) charges per fan for various values of J.

Eight attitudinal patterns emerged within the frame of the experiment. Taking into account the internal personal constraint J, of the opinion (O), emotional (E) and action (A) components, those patterns could be coded as: {7.5/7.5/7.5} (the most extreme, i.e. potentially violent sub-sample),{7.5/7.5/5.5},{7.5/7.5/4.5},{7.5/7.5/3.5},{7.5/6.5/7.}, {7.5/6.5/5.5},{7.5/6.5/4.5} and {7.5/6.5/3.5} (the least

extreme subsample). For the semantic content of those patterns, see the questionnaires given in the former chapter.

## Principal component analysis

Although the Guttman-Kaiser criterion tends to maximize the number of significant principal components with eigenvalues larger than 1, the analysis revealed one general component that explained 98% of the total variance with eigenvalue $\lambda=6.95$. All communalities were highly significant $0.998\pm0.004$ as well as the loadings of the variables on the principal component $0.987\pm0.003$.

## Discussion

The experimental evidence supports the basic predictions of the analytical model (Hristovski, 1997). One has to notice that attitudinal layers of coherence in the experiment do not emerge from local (nearest neighbour) or any other distributed form of interactions among the individuals, which has been accepted as one of the main concepts in social modeling (e.g. Nowak et al., 1990; Latane, 1996; Vallacher & Jackson, 2009). Within the framework of this model they arise as a consequence of commonalities in their personal constraints (common identification with their favorite team), which form their common mode of coupling with the external stimuli, i.e. referee's decisions). In the experiment, that is the reason why, under the influence of environmental stimuli, the primordial symmetry of the homogenous state was broken (rearranged) and the creation of individual as well as collective attitudes emerged.

Polarized states within the observed sample of fans were not attained due to the semantic specificities of the stimuli given in the experiment. If you recall, the referee was making incorrect decisions concerning their favorite team. The interaction of the incorrect decisions and the common identification with the favorite team led to a strong unidirectional coupling of the whole sample (N = 97) with external events. The condition that is necessary to cause a polarized attitude states is the existence of a different coupling sign with the external events of the social ensemble. That was not the case in this sample, but it is highly feasible to fulfill these conditions using a certain type of experimental design.

Thus, the evidence corroborates the model's prediction which claims that: when locally non - interacting individuals have mutually strongly correlated couplings concerning the social environment, the individual attitudes are created as a consequence of the symmetry breaking mechanism, which notwithstanding the inter-individual differences, for a finite number of semantically equivalent or similar stimuli, leads to the state of an attitudinal coherence.

The ensemble susceptibility and fluctuations showed a clear critical behavior. Shortly after the onset of environmental events they enhanced which signaled a destabilization of the previous state and a transition to a coherent attitude formation. Both measures attained maximal values for B = 2 - 4. This means that under global coupling and referee's decisions that fans perceive as moderately offensive, the maximal responsiveness of fan groups and their variability in attitudes has to be expected within this range of number of events. This critical behavior is a hallmark of the phase transition phenomena, in which the system attains order through fluctuations of the collective behavior. In fact, it is the non-local coupling among the fans that breaks the symmetry under the impinging referee decisions. In other words, the common unidirectional non-local coupling among fans for small B values competes with the inter-individual differences in other psychological degrees of freedom. That is why we see a unidirectional enhancement of fluctuations. All fans react in the same direction (i.e. they brake the initial symmetry present at B=0), but they, as individuals, differ in other sets of personal constraints and that contributes to the early 'disagreement', i.e. the enhancement of fluctuations. Such unidirectional fluctuations actually push the expected <U> towards the final coherence value, which emerges for higher B. For higher B values the common identification wins, the fluctuations are suppressed and coherence emerges. In this case, the referee's decisions have the role of a control parameter which destabilizes the neutral attitudinal state present at B=0, and through fluctuations directs the system into a coherent state. Such coherence was obvious in the principal component analysis results where seven variables were reduced to only one collective variable (order parameter) accounting for more than 98% of the total variance. As the system, under the influence of the environmental events, approached the transition point, fluctuations of this collective variable grew and afterwards were quickly suppressed as the social ensemble converged to the new coherent state. This behavior represents a strong evidence of the highly integrated nature of opinion, emotional and action spaces on individual and social level. It is also interesting to pay attention to the fact that largest coherence was attained in the opinion space whereas the emotional and particularly the action space were more segregated. This fact may be a result of the more refined structure of the fan sample concerning their responses to external events.

In the experiment, the existence of equifinality, (the property of the social system as well as of its elements (individuals) to attain the same final state, starting from various initial conditions and by using various paths), was demonstrated. One can consider that the variability of paths is an effect of the impact of a large number of weakly correlated intrapersonal degrees of freedom on the evolution of individual attitudes. In this investigation we assumed that this is the generator responsible for the presence of attitude variability (fluctuations) over a large interval of environmental stimuli (see Figure 6). It can be seen that for larger values of B (B ≥ 3 ÷ 4) the attitudinal variability weakened as the common coupling with the environmental events became dominant and suppressed the personal degrees of freedom, so that the strongly correlated individual attitudes were, literally, sucked into a single collective attitude. Thus, the process of creation and evolution of the attitudes within this model shows robust properties concerning the initial conditions (i.e. equifinality). Moreover, the high goodness of fit for different number of participants (see Figures 2, 3 and 4) supports the robust properties of this model in respect to the cardinality of the observed social system as well.

Furthermore, layers of coherence emerged as an effect of the existence of intrapersonal constraint J, that represented an extremity of individual responses in the cognitive, emotional or action space. This constraint introduced a potential for multistability into the system, that is, a large number of stable collective coherent attitudes (attractors) were created by the social system under the impact of external stimuli. In other words the number J taxonomizes the initially homogenous sample into attitudinal subsamples (stratums). This leads to a notion of a non - local coherent layering (stratification). That is in fact the phenomenon of the capability of a certain number of spatially randomly distributed and mutually non - interacting individuals to attain different layers (stratums) of attitudinal coherence, after being exposed to equal environmental stimulation for a certain amount of time.

In reality, at least some of the initial attitudinal states, before they become subjected to change by the inter-individual interactions, are probably spatially randomly distributed on the basis of this effect. According to this, for instance, the patterns {7.5/7.5/7.5} and {7.5/6.5/7.5} could be considered as a formation of potentially violent proto-groups, that afterwards could attain (by local network interactions) all the properties, characteristic for a violent mob, among including spatial clustering and collective violent action. Another important theoretical and practical aspect stems from these findings. The emergence of coherence as a result of suppression of individual degrees of freedom leads to a possibility of efficient control and manipulation of such social ensembles by a third person or institution because as the majority of personal differences are suppressed, the whole ensemble may be treated as a single collective entity controlled by a common coherent information. Hence, instead of dealing with a multitude of individual degrees of freedom, i.e. opinions, emotions and action preferences, one actually has to deal with one collective coherent person. On the other hand, the pattern {7.5/7.5/3.5/} implies a social proto-group, that has a potential to evolve in a well formed group trying to realize its goals through institutional procedures ($J = 3.5$).

The combination of the results of this model with those obtained by local models, can lead to considering the predicted final states within this model as an initial state of a model with local interactions among the elements. Vice versa, the coherent clusters (coherent pockets of attitude holders) obtained in local models could be treated as a common pool of information on whose basis another set of attitudes and non - local coherence or polarization in the social system could emerge as a consequence of the symmetry breaking under new and different environmental stimuli. It would be particularly interesting to investigate the possibility of change of personal constraints J as a consequence of local interactions among fans. The possibility that the J constraint is subject to change under local interactions, such as peer pressure, is quite real. In that case, individual fans possessing different J constraints, under conditions of global coupling with environmental events, may change their personal constraints when subject to local interactions.

The way of forming groups with common personal constraints, from proto-groups, might be an interesting topic for future research. How the interpersonal interactions grow depending on the common views of environmental events and their action preferences? This problem is obviously fundamental to group or even team formation process, where one may investigate the formation of network connections among the elements of the proto-group. Thus, the iterations of attitude creation under external stimulation and attitude change by local interactions (iterations of the external and the internal social impact) might enable the existence of temporal interchange of the roles among the constraints (non - specific control parameters or boundary conditions) and order parameters of the social system, which is one of the prominent properties of evolving systems.

Yet another interesting direction of future investigation can be inferred from the results of this investigation: The least violent coherent layer in the action space $J = 3.5$ ($<U> = 4$) was formed by 79% of the total number of female fans. Hence a further issue emerges: How J constraints depend on other social constraints (e.g. the influence of family attitudes toward female and male populations), as well as on possible influences of some gender biochemical differences, e.g. the level of the testosterone?

Furthermore, it would be interesting to treat the behavior of social ensembles when an environmental control parameter is not solely the accumulating number of referee's decisions B, but also their time characteristics like their frequency in time. It may happen that the coherence attained in some of the investigated spaces within this study, e.g. the emotional and action space, declines, as a result of declining frequency of environmental stimulation and yet in the space of opinions the coherence may remain unchanged. This may point to differences in the dependence on the opinion space on one side, and the emotional and action spaces on the other. The behavior of social ensembles under varying neutral, positive and negative referee's decisions is yet another direction of future investigations which may reveal somewhat different results than those obtained in this research.

### Acknowledgements

We would like to thank anonymous reviewers for their useful comments and suggestions on the earlier draft of this paper.

### References

Cox, J. T. & Durrett, R. (1991). Nonlinear voter models. In: Random Walks, R. Durrett & H. Kesten (Eds.). *Brownian Motion and Interacting Particle Systems* (pp. 189 – 202). Boston: Birkhauser.

Dyke, C. (1988). Entropy, Information and Evolution. In B. Weber, D. Depew & J. Smith (eds) (p. 355). Cambridge: MIT Press.

Gilpin, M. E. (1975). *Limit cycles in competition communities. Am. Nat.*, 108, 207 - 228.

Hare, A.P., (1962). *Handbook of small group research*. New York: Macmillan Publishers.

Hristovski, R. (1997). Condensation of coherent states in social groups by the symmetry breaking mechanism. *Proceedings of the 6-th International Congress on FIS Communications*, Nish, p. 279 - 282.

Isnard, C. A., Zeeman, E. C. (1976). Some models of catastrophe theory in the social sciences. In I. Collins (ed): *Use of Models in the Social Sciences* (pp. 44-100). London: Tavistock.

Kondratjev, A.S. and Romanov, V.P. (1992). *Problems in statistical physics. (Zadaci po statisticeskoi fizike)*. Moskva: Nauka.

König, R. (1973). *Basic Methods and Technics in Sociological research. (Grundlegende Methoden und Techniken der empirischen Socialforschung)*. 1° Teil, Band 2, 3° Auflage. Stuttgart: Enke Verlag.

Latane, B. (1996). Dynamic social impact: Robust predictions from simple theory. In R. Hegselman, U. Mueler, and K. Troitzsch, (Eds.): *Modelling and Simulating in the Social Sciences from a Philosophy of Science point of view*. Dordrecht: Kluver.

Lazlo, E. (1985).. In I. Prigogine and M. Sanglier (eds): *The Laws of Nature and Human Conduct (p. 298)*. Brussels: Task Force of Research Information and Study on Science.

Liggett, T. M. (1994). Coexistence in threshold voter models. *Ann. Prob*. 22, 764 - 802.

May, R. M. and Leonard, W. L. (1975). Nonlinear aspects of competition between three species. *SIAM J. Appl. Math*. 29, 243 - 253.

Nowak, A., Szamrej, J. and Latane, B. (1990): From private attitude to public opinion: A dynamic theory of social impact. *Psychological Review*, 97, 362 - 367.

Nowak, A. and Hegselmann, R. (1993). Emergence of public opinion, the role of individual differences and geometry of social space. In R. Hegselmann (ed.), *Chaos and Order in Nature and Society*. Viena: Hölder-Pichler.

Nowak, A. and Latane, B. (1994). Simulating the emergence of social order from individual behavior. In: N. Gilbert & J. Doran (Eds.), *Simulating societies: The computer simulation of social processes*. London: University College London Press.

Nowak, A., and Vallacher, R. R. (1998): Toward computational social psychology: Cellular automata and neural network models of interpersonal Dynamics. In S. J. Reed & L. C. Miller (Eds.) (pp. 277 – 311), *Connectionist models of social reasoning and social behavior*. Mahwah, NJ: Erlbaum.

Nowak, A., Vallacher, R. R. and Burnstein, E. (1998). Computational social psychology: A neural network approach to interpersonal dynamics. In W. Liebrand, A. Nowak & R. Hegselmann (Eds.) (pp. 97 – 125), *Computer modeling and the study of dynamic social processes*. New York: Sage.

Renfrew, A. C. and Poston, T. (1978). Discontinuities in the endogenous change of settlement pattern. In K.L. Cooke, & C. A. Renfrew: *Transformations: Mathematical Approaches to Culture Change* (eds.), New York: Academic Press.

Rumelhart, D. E. and McCleland, J. L. (1986). *Parallel distributed Processing: Explorations in the microstructure of cognition.* Vol. 1. Foundations. Cambridge, MA: MIT Press/ Bradford.

Shelling, T. (1969). *Models of segregation. Am. Econom. Rev*., 59, 488 - 493.

Shelling, T. (1971). *Dynamic models of segregation. J. Math. Sociol.*, 1, 143 - 186.

Silvertown, J., Holtier, S., Johnson, J. and Dale, P. (1992). *Cellular automaton models of interspecific competition for space - the effect of pattern on process. J. Ecol*. 80, 527 - 534.

Tainaka, K. (1995): Indirect effect in cyclic voter model models. *Phys. Lett. A*, 207, 53 - 57.

Vallacher, R. R., and Jackson, D. (2009). Thinking inside the box: Dynamical constraints on mind and action. *European Journal of Social Psychology*, 39, 1226-1229.

Vallacher, R. R., Coleman, P. T., Nowak, A., and Bui-Wrzosinska, L. (2010). Rethinking intractable conflict: The perspective of dynamical systems. *American Psychologist*, 65, 262-278.

Weidlich, W. (1978). Dynamics of interacting social groups. In: H. Haken (Ed.) (pp. 269 – 282), Cooperative Effects. Progress in synergetics. Amsterdam: North Holland.

Zeeman, E. C., Hall, C., Harrison, P. J., Marriage, H. and Shapland, P. (1976). A model for institutional disturbances. *Br. J. Math. Statist. Psych*. 29, 66 - 80.